# Direct evidence: twisted flux tube emergence creates solar active regions

MacTaggart D.[1], Prior C.[2], Raphaldini B.[2], Romano, P.[3], Guglielmino, S.L.[3]

[1]School of Mathematics and Statistics, University of Glasgow, Glasgow, G12 8QQ, UK

[2]Department of Mathematical Sciences, Durham University, Durham, DH1 3LE, UK

[3]INAF-Osservatorio Astrofisico di Catania, Via S. Sofia 78, I-95123 Catania, Italy

**The magnetic nature of the formation of solar active regions lies at the heart of understanding solar activity and, in particular, solar eruptions. A widespread model, used in many theoretical studies, simulations and the interpretation of observations[1,2,3,4], is that the basic structure of an active region is created by the emergence of a large tube of pre-twisted magnetic field. Despite plausible reasons and the availability of various proxies suggesting the veracity of this model[5,6,7], there has not yet been any direct observational evidence of the emergence of large twisted magnetic flux tubes. Thus, the fundamental question, "are active regions formed by large twisted flux tubes?" has remained open. In this work, we answer this question in the affirmative and provide direct evidence to support this. We do this by investigating a robust topological quantity, called *magnetic winding*[8], in solar observations. This quantity, combined with other signatures that are currently available, provides the first direct evidence that large twisted flux tubes do emerge to create active regions.**

Twisted flux tubes are prominent candidates for the progenitors of solar active regions. Twist allows a flux tube to suffer less deformation in the convection zone compared to untwisted tubes, thus allowing it to survive and reach the photosphere to emerge[2,9,10]. Further, simulations of twisted flux tube emergence have reproduced signatures that can be found in observations[1,3]. This has led to observational proxies (such as sigmoidal field lines in the atmosphere[4] and "magnetic tongue" patterns in magnetograms[6,7]) that are indicative of twisted tube emergence. Although highly suggestive, such proxies cannot provide direct and conclusive evidence of the emergence of a twisted flux tube from the convection zone to the solar atmosphere. This is because their signatures can also be created by magnetic fields that are not pre-twisted flux tubes[11]. Further, the signatures of these proxies can also be created by photospheric motions deforming simple magnetic fields, e.g. shearing flows along the polarity inversion line of an active region leading to sigmoidal field lines[12,13].

Twist in a magnetic field is a manifestation of *magnetic topology*, which describes the connectivity of field lines. A classical measure of magnetic topology is *magnetic helicity*[14]. This quantity describes the field line topology (in terms of Gauss linkage[14] or winding[8], depending on the precise application) weighted by magnetic flux. Further, the flux of magnetic helicity through the photosphere can be measured in solar observations[15,16], so this topological quantity can potentially indicate what kind of magnetic topology is emerging. Many works[17-24] have studied the injection of this quantity in active regions, but a clear-cut indication of an emerging magnetic field's magnetic topology from magnetic helicity flux has remained elusive. One reason for this is that magnetic helicity combines two distinct properties, field line topology and magnetic flux, into one quantity. This combination can result in confounding the interpretation of the helicity flux. For example, an emerging field with simple (weak) field line topology could have a very strong field strength. Thus, a strong input of helicity does not necessarily indicate that complex (strong) magnetic topology is emerging into the solar atmosphere. Similarly, a magnetic field with a highly twisted topology but weak flux, can result in a weak helicity output, and so also not give an accurate indication of the true nature of the field line topology emerging into the atmosphere.

*Magnetic winding*[8] is a renormalization of magnetic helicity that removes the magnetic flux weighting, and thus provides a direct measure of magnetic topology. Despite its close connection to magnetic helicity, magnetic winding can behave very differently in an evolving magnetic field and, hence, provide new and distinct information. Further, magnetic winding flux can be calculated from observations just like the helicity flux. Magnetic winding flux has been studied in simulations of magnetic flux emergence, including twisted flux tubes and more complex magnetic topologies[25,26]. Over a wide range of physical parameters, the accumulation of winding (the time-integrated winding flux) during the initial emergence of a twisted flux tube produces a consistent signature: an initial increase in the magnitude of the winding input followed by a plateau. This signature indicates that the twisted tube (or, at least, a substantial part of it) passes through the photospheric boundary (where the flux calculation is performed) and then essentially remains above this plane afterwards (until perhaps much later times when the active region begins to decay, but we are only interested in the initial stages of emergence here). Even the effects of convection, in simulations, which act to drag down parts of emerged field and create a "serpentine" structure, have little effect on the winding input signature, as the magnetic topology is still dominated by the twist in the tube that remains primarily above the photospheric boundary. By contrast, the net magnetic helicity input can change sign during the emergence of twisted flux ropes due to the strength of convective down-flows, leading to a potential misinterpretation of the magnetic field's structure, i.e. a failure to diagnose the true twisted nature of the emerging field[26]. An example of the winding signature described above, from a magnetohydrodynamic simulation of the initial stages of the emergence of a twisted flux tube, is shown in Figure 1.

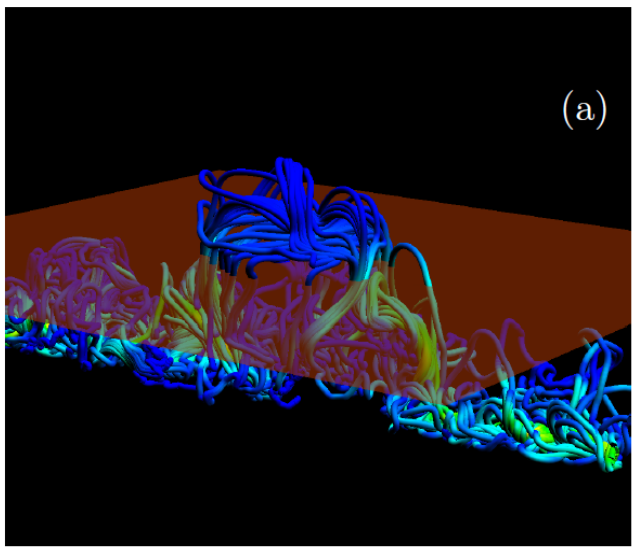
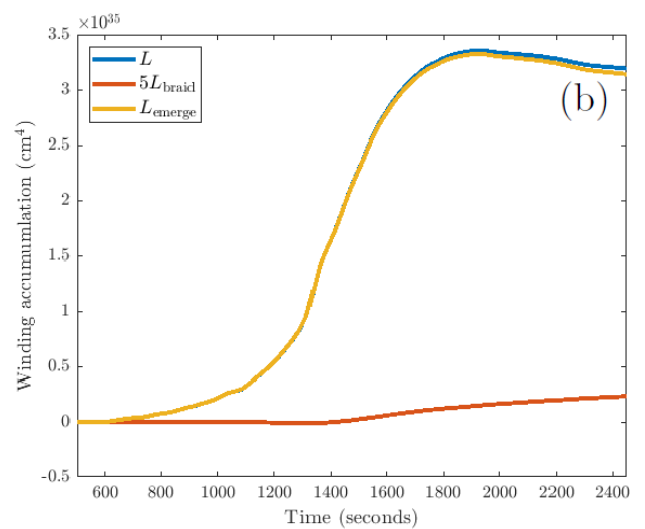

**Figure 1: simulation of the initial emergence of a twisted magnetic tube.** (a) displays field lines at $t$=2500 seconds, with a slice indicating the photospheric boundary. Darker colours (mainly blue) indicate weaker field strengths and lighter colours (green and yellow) indicate stronger field strengths. In (b), the emergence accumulation $L_{emerge}$, the braiding accumulation $L_{braid}$, and the total winding accumulation $L = L_{emerge} + L_{braid}$ are displayed. $5 \times L_{braid}$ is displayed on the figure in order to convey clearly how $L_{braid}$ develops in time.

Figure 1(a) shows the emerged flux tube, at $t$=2500 seconds, which has been deformed significantly by convection and developed a "serpentine" field line structure. Any clear visual resemblance to a twisted tube is lost and the emerged magnetic field has the form of a collection of magnetic arcades. Indeed, it is important to point out that when we refer to twisted tube emergence, we are not implying that a coherent twisted flux tube emerges bodily from the convection zone to the corona. Instead, the tube emerges into the photosphere and the bulk of the twist remains there. As the magnetic field expands into the atmosphere, it evolves into one or several shear arcades, depending on the complexity of the region and the subsequent magnetic reconfiguration[1,2,3].

Figure 1(b) shows the accumulation of winding above the photospheric boundary. There are three profiles to consider: the total winding accumulation, the braiding accumulation and the emergence accumulation. The calculation of the magnetic winding depends on two components of the magnetic field line velocity $\boldsymbol{u}$ on the planar photospheric boundary: a velocity due to in-plane motion, $\boldsymbol{v}_{||}$, and a velocity due to the emergence of the magnetic field, $-v_z \boldsymbol{B}_{||}/B_z$, where $\boldsymbol{v} = (\boldsymbol{v}_{||}, v_z)$ is the plasma velocity field and $\boldsymbol{B} = (\boldsymbol{B}_{||}, B_z)$ is the magnetic field (parallel subscripts indicate "parallel to the photospheric plane"). The braiding accumulation $L_{braid}$ describes the winding input due to only the in-plane velocity and measures the entanglement of the magnetic field due to horizontal photospheric motions. The emergence accumulation $L_{emerge}$ describes the winding input due to only the emergence velocity and measures the contribution to winding due to pre-entangled emerging magnetic field (see Methods section for more details). The total winding accumulation follows the signature described previously. Importantly, the emergence accumulation dominates strongly over the braiding accumulation. This result states that the input of magnetic winding into the atmosphere is due primarily to the emergence of a pre-entangled magnetic field and not due to horizontal motions twisting the magnetic field at the photosphere. The signature in Figure 1(b) indicates that the emerging entangled field is dominated by a positive twist of the magnetic field lines. More complex field line topologies would not create such a simple accumulation signature[25]. The magnetic winding is robust enough to detect the twisted field line topology despite the substantial deformation to the original flux tube in the convection zone.

We now present direct evidence of twisted flux tube emergence in solar observations. For the purpose of this article, we focus on "clean" cases (isolated and coherent bipolar regions) that can be compared to, as closely as possible, the results from simulations. The active regions in the observations we consider are much larger than that of the simulation presented earlier. However, the result that we are testing for is not strictly scale-dependent, i.e. the same signature applies to large or small tubes, as long as they emerge above the photospheric boundary.

We first consider the National Oceanic and Atmospheric Region (NOAA) active region AR11318. This region has been studied in detail by measuring a range of different observational signatures[23]. These signatures, however, are still not enough to confirm that a coherent twisted flux tube has emerged to create this bipolar region. Figure 2 shows the winding accumulation for the initial emergence of AR11318, starting from 19:12UT on October 11th 2011 and lasting for 80 hours. To perform winding flux calculations, we use *S*pace-Weather *H*elioseismic and Magnetic Imager (HMI) *A*ctive *R*egion *P*atches (SHARP)[27] vector magnetograms taken by HMI on board of the Solar Dynamics Observatory

(SDO), and we determine the plasma velocity using the Differential Affine Velocity Estimator for Vector Magnetograms (DAVE4VM)[28]. The SHARP vector magnetograms have a pixel resolution of 0''.5 and a time cadence of 720s.

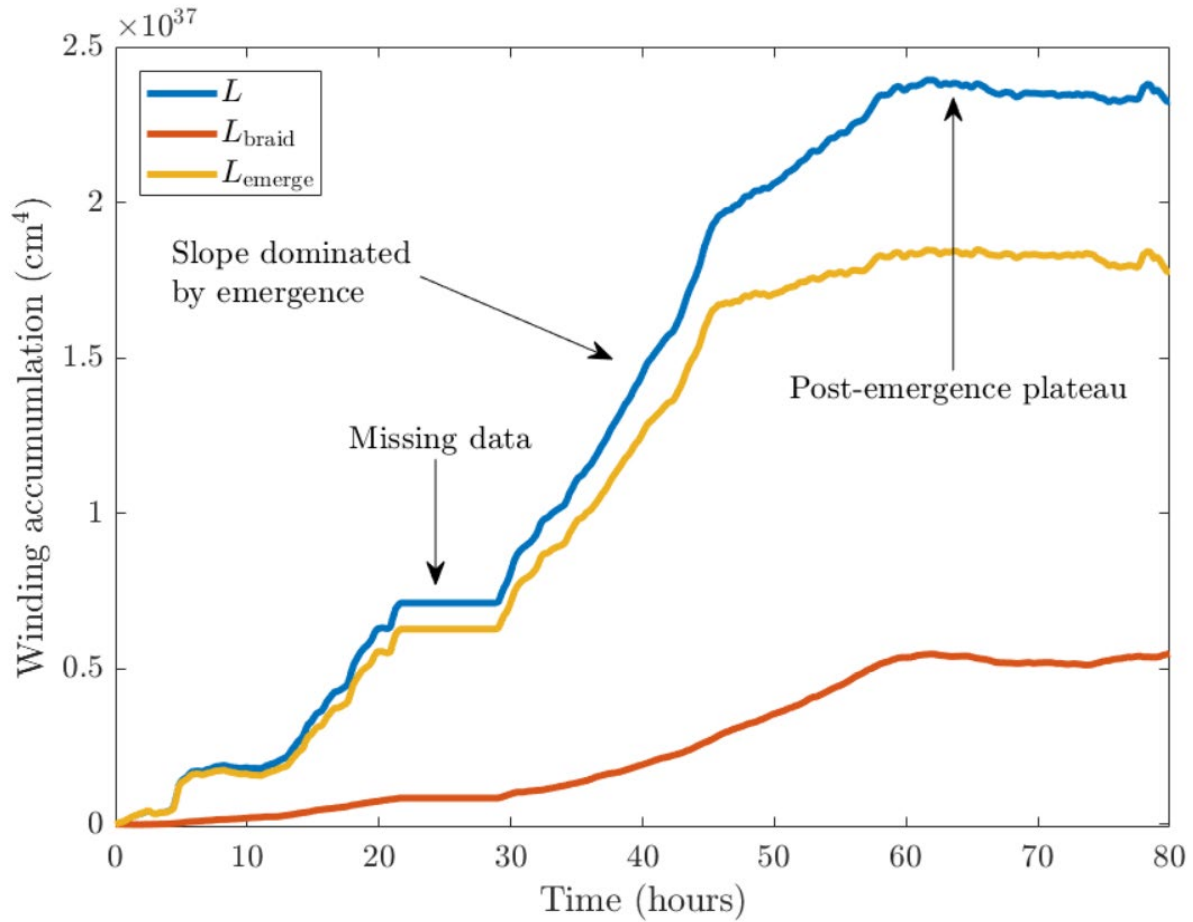

**Figure 2: Winding accumulation for AR11318.** The emergence accumulation $L_{emerge}$, the braiding accumulation $L_{braid}$, and the total winding accumulation $L = L_{emerge} + L_{braid}$ are displayed. The missing data are between *t*=20 and *t*=30 hours.

Between *t*=20 and *t*=30 hours there are missing data which result in an artificial plateau in the accumulation curves. Ignoring this data gap, the total winding accumulation $L$ (calculated with both velocity components) follows the signature of Figure 1(b) (and other simulations[25,26]), namely a strong rise followed by a plateau. It is clear from Figure 2 that the emergence accumulation dominates strongly over the braiding accumulation, and so the winding input is due primarily to the emergence of a pre-twisted structure rather than an untwisted structure whose twist develops in the solar atmosphere due to photospheric motions. If, during the data gap, the curves were to follow their approximately constant gradients (a reasonable assumption based on the behaviour of the gradients immediately before and after the data gap), the difference between $L_{emerge}$ and $L_{braid}$ would be even greater than that displayed in Figure 2. This signature, together with those found previously[23] for AR11318 (magnetic tongues and the development of sigmoidal field lines) that follow the emergence evolution expected from simulations of twisted tube emergence[1,2,3], clearly and directly demonstrates that the region was formed by an emerging twisted flux tube.

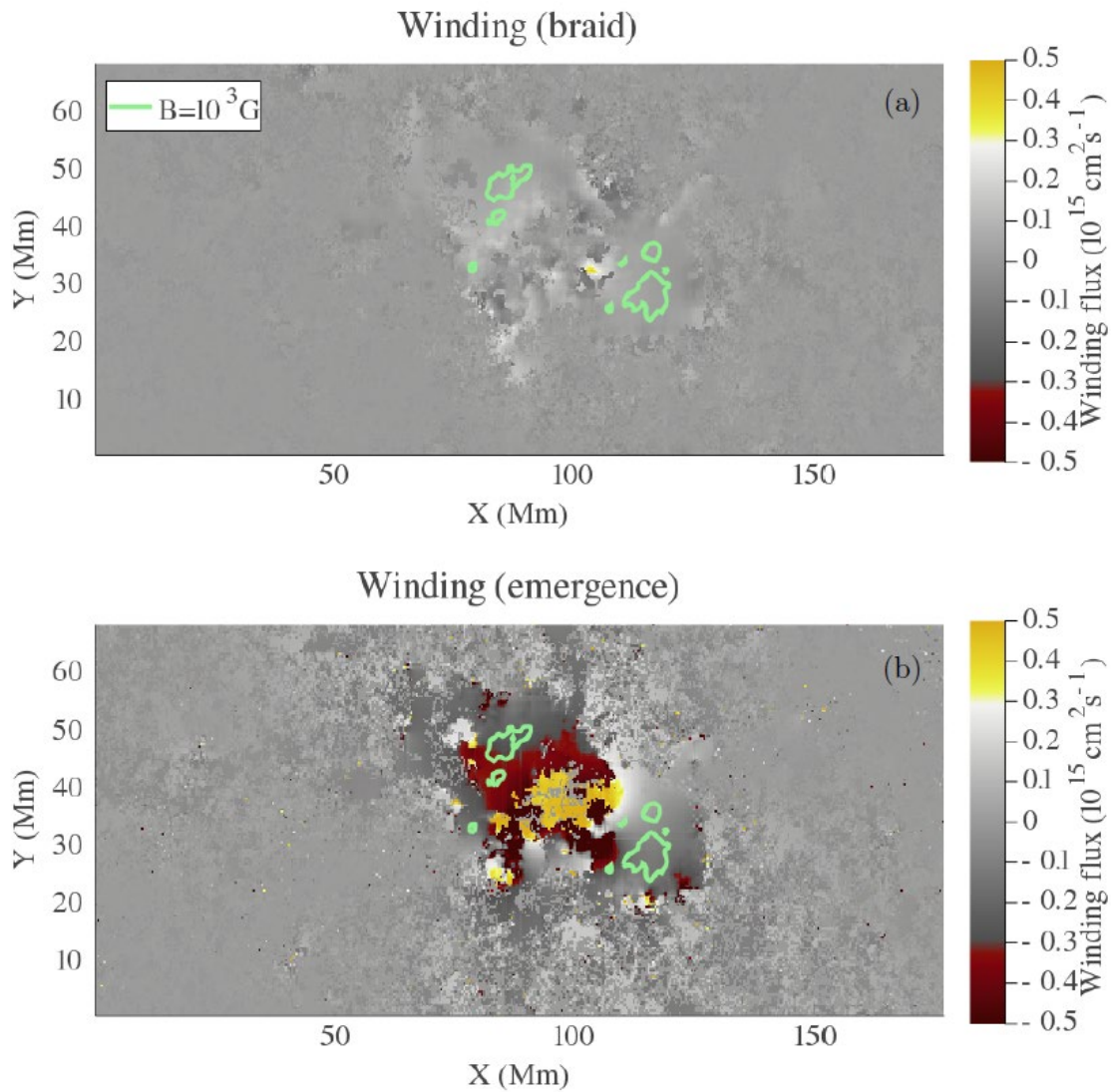

**Figure 3: Winding flux distributions.** Distributions of (a) $dL_{braid}/dt$ and (b) $dL_{emerge}/dt$ of AR11318

at $t = 42$ hours. The green contours indicate a field strength of $B_z = 1$kG (and thus indicate the sunspot locations).

Figure 3 displays distributions of (a) $dL_{braid}/dt$ and (b) $dL_{emerge}/dt$ of AR11318 at $t = 42$ hours. The contribution due to emergence dominates the region between the sunspots, where the horizontal part of the twisted tube emerges. The contribution due to braiding is much weaker, apart from a few isolated locations, and spread throughout the domain. There are also coherent patches at the sunspots which move apart during the emergence of the region.

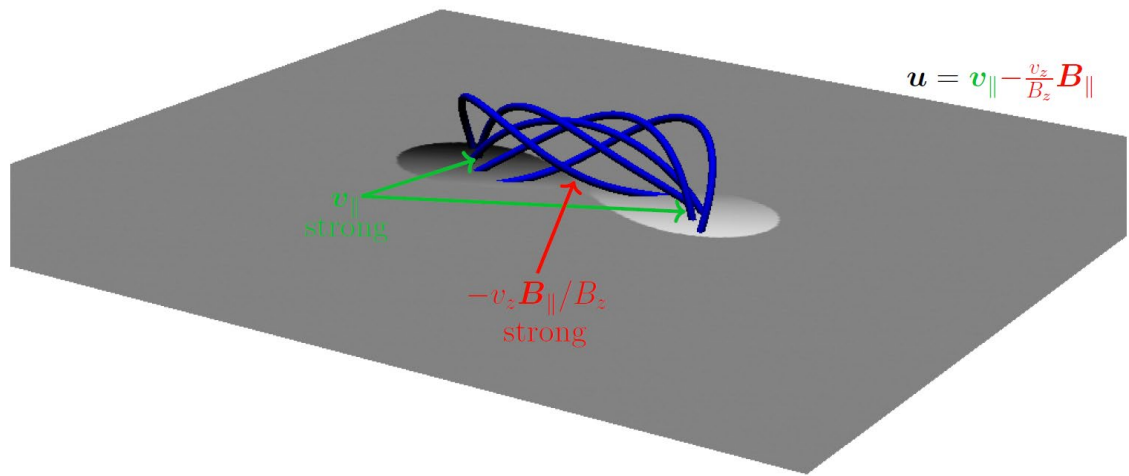

**Figure 4: Influence of velocity components.** Field lines of an emerging twisted flux tube are shown above a planar boundary representing the photosphere. A map of $B_z$ is displayed on the photospheric boundary. The field line velocity $\boldsymbol{u}$ has two components: in-plane (green) and emergence (red), which play important roles at different parts of an emerging twisted flux tube, as indicated.

The behaviour of the winding profiles and distributions can be explained with the aid of the cartoon shown in Figure 4. This cartoon indicates what parts of an emerging twisted magnetic field contribute to the winding accumulation. The emergence velocity is most prominent between the sunspots where $B_z$ is weaker and $\boldsymbol{B}_{\parallel}$ is at its strongest. It is between the sunspots where the main bulk of the twisted tube emerges and this results in a strong signal in the emergence accumulation (see Figure 3(b)). As mentioned before, magnetic topologies different from twisted flux tubes can also be detected by the winding accumulation, but they have more complex accumulation profiles than that for twisted flux tubes[25]. The in-plane velocity is stronger near the sunspots and is generally weaker between the sunspots. These considerations are important for understanding why the magnetic winding flux can detect the emergence of large-scale entanglement (twist in this case) clearly whereas the magnetic helicity flux may not. In magnetic helicity flux calculations, each point on the photosphere is weighted by $B_z$. The effect of this is to make the contribution of the emergence velocity small, since $B_z$ is weak between the sunspots (for the simple emerging twisted tube under consideration here). Hence, the magnetic helicity flux input is dominated by the in-plane velocity contribution at the sunspots, which is weighted by strong $B_z$. Therefore, the magnetic helicity flux may not adequately diagnose the emergence of twisted magnetic field emerging between the sunspots. Extended data figure 1 displays the helicity accumulation for AR11318. The importance of the emergence and braid components is now the opposite to that of the winding, in line with our description above. Extended data figure 2 shows the helicity input distributions at $t = 42$ hours. Again, as per the above discussion, the helicity input is dominated by the sunspots, due to the weighting of strong $B_z$, and so the bulk of the signature of the twisted flux tube emerging between the spots is missed.

Other "clean" examples of twisted flux tube emergence can be found. Extended data figure 3 shows the winding accumulation during the initial emergence of AR12203, starting at 13:00UT on 30th October 2014 and lasting for 120 hours. In this region, negative (left-handed) twist emerges into the atmosphere. Again, the emergence accumulation dominates strongly over the braiding accumulation and provides a clear signature that the magnetic topology of this region comes initially from emergence and not photospheric shearing motions. Examining other signatures, such as those studied previously for AR11318[23], reveals a similar initial evolution. Extended data figure 4 displays the helicity accumulation of AR12203. As for AR11318, the importance of the emergence and braiding contributions has now switched compared to the winding accumulation. These helicity inputs, unlike the winding inputs, do not give a clear indication of how the topology of the emerging magnetic field enters into the atmosphere.

Magnetic winding, in combination with other measurable quantities (such as magnetic helicity) and proxies, provides a powerful analysis tool that can give direct information about magnetic topology. We have given examples of active region observations where the magnetic winding gives a clear indication that the emerging magnetic field is composed of pre-twisted magnetic field. This confirms that, as assumed in many theoretical studies, pre-twisted flux tubes play a fundamental role in active region formation. The pre-twisted magnetic field that emerges into the photosphere represents a source to explain, self-consistently, the shearing, rotational and compressional motions invoked in models of coronal mass ejection formation. These motions can develop due to the transportation of twist into the higher atmosphere as the magnetic field expands into the corona[1,3,29,30]. Although we have presented evidence that twisted flux tubes can create active regions, it is likely that other magnetic topologies also emerge to create active regions. This is an important area of research, for which magnetic winding will play a pivotal role.

1. Cheung, M.C.M. & Isobe, H. Flux Emergence (Theory). *Living Rev. Sol. Phys.* **11**, 3 (2014)

2. Fan, Y. Magnetic Fields in the Solar Convection Zone. *Living Rev. Sol. Phys.* **6**, 4 (2009)

3. Hood, A.W., Archontis, V. & MacTaggart, D. 3D Magnetic Flux Emergence Experiments: Idealized Models and Coronal Interaction. *Sol. Phys.* **278**, 3-31 (2012)

4. van Driel-Gesztelyi, L. & Green, L. Evolution of Active Regions. *Living Rev. Sol. Phys.* **12**, 1 (2015)

5. Archontis, V., Hood. A.W., Savcheva, A., Golub, L. & Deluca, E. On the Structure and Evolution of Complexity in Sigmoids: A Flux Emergence Model. *Astrophys. J.* **691**, 1276-1291 (2009)

6. Luoni, M.L., Démoulin, P., Mandrini, C.H. & van Driel-Gesztelyi, L. Twisted Flux Tube Emergence Evidenced in Longitudinal Magnetograms: Magnetic Tongues. *Sol. Phys.* **270**, 45-74 (2011)

7. Poisson, M., Mandrini, C.H., Démoulin, P. & López Fuentes, M. Evidence of Twisted Flux-Tube Emergence in Active Regions. *Sol. Phys.* **290**, 727-751 (2015)

8. Prior, C. & MacTaggart, D. Magnetic winding: what is it and what is it good for? *Proc. R. Soc. A* **476**, 20200483 (2020)

9. Emonet, T. & Moreno-Insertis, F. The Physics of Twisted Magnetic Flux Tubes Rising in a Stratified Medium: Two-dimensional Results. *Astrophys. J.* **492**, 804-821 (1998)

10. Martínez-Sykora, J., Moreno-Insertis, F. & Cheung, M.C.M. Multi-Parametric Study of Rising 3D Buoyant Flux Tubes in an Adiabatic Stratification using AMR. *Astrophys. J.* **814**, 2 (2015)

11. Prior, C. & MacTaggart, D. The emergence of braided magnetic fields. *Geophys. Astrophys. Fluid Dyn.* **110**, 432-457 (2016)

12. Amari, T., Luciani, J.F., Aly, J.J., Mikic, Z. & Linker, J. Coronal Mass Ejection: Initiation, Magnetic Helicity, and Flux Ropes, I. Boundary Motion-Driven Evolution. *Astrophys. J.* **585**, 1073-1086 (2003)

13. Aulanier, G., Janvier, M. & Schmieder, B. The standard flare model in three dimensions. I. Strong-to-weak shear transition in post-flare loops. *Astron. Astrophys.* **543**, A110 (2012)

14. Moffatt, H.K. The degree of knottedness of tangled vortex lines. *J. Fluid Mech.* **35**, 117-129 (1969)

15. Démoulin, P. & Berger, M.A. Magnetic Energy and Helicity Fluxes at the Photospheric Level. *Sol. Phys.* **215**, 203-215 (2003)

16. Pariat, E., Démoulin, P. & Berger, M.A. Photospheric flux density of magnetic helicity. *Astron. Astrophys.* **439**, 1191-1203 (2005)

17. Yamamoto, T.T., Kusano, K., Maeshiro, T., Yokoyama, T. & Sakurai, T. Magnetic Helicity Injection and Sigmoidal Coronal Loops. *Astrophys. J.* **624**, 1072-1079 (2005)

18. Jeong, H. & Chae, J. Magnetic Helicity Injection in Active Regions. *Astrophys. J.* **671**, 1022-1033 (2007)

19. Yang, S., Zhang, H. & Büchner. J. Magnetic helicity accumulation and tilt angle evolution of newly emerging active regions. *Astron. Astrophys.*, 333-340 (2009)

20. Chandra, R., Pariat, E., Schmieder, B., Mandrini, C.H. & Uddin, W. How Can a Negative Magnetic Helicity Active Region Generate a Positive Helicity Magnetic Cloud? *Sol. Phys.* **261**, 127-148 (2010)

21. Liu, Y. & Schuck, P.W. Magnetic Energy and Helicity in Two Emerging Active Regions in the Sun. *Astrophys. J.* **761**, 105 (2012)

22. Park, S.-H. et al. The Occurrence and Speed of CMEs Related to Two Characteristic Evolution Patterns of Helicity Injection in their Solar Source Regions. *Astrophys. J.* **750**, 48 (2012)

23. Romano, P., Zuccarello, F.P., Guglielmino, S.L. & Zuccarello, F. Evolution of the Magnetic Helicity Flux During the Formation and Eruption of Flux Ropes. *Astrophys. J.* **794**, 118 (2014)

24. Vemareddy, P. & Démoulin, P. Successive injection of opposite magnetic helicity in solar active region NOAA 11928. *Astron. Astrophys.* **597**, A104 (2017)

25. Prior, C. & MacTaggart, D. Interpreting magnetic helicity flux in solar flux emergence. *J. Plasma Phys.* **84**, 775850201 (2019)

26. MacTaggart, D. & Prior, C. Helicity and winding fluxes as indicators of twisted flux emergence. *Geophys. Astrophys. Fluid Dyn.* **115**, 85-124 (2021)

27. Hoeksema, J.T. et al. The Helioseismic and Magnetic Imager (HMI) Vector Magnetic Field Pipeline: Overview and Performance. *Sol. Phys.* **289**, 3483-3530 (2014)

28. Schuck, P.W. Tracking Vector Magnetograms with the Magnetic Induction Equation. *Astrophys. J.* **683**, 1134-1152 (2008)

29. Magara, T. & Longcope, D.W. Sigmoid Structure of an Emerging Flux Tube. *Astrophys. J.* **559**, L55-L59 (2001)

30. Fan, Y. The Emergence of a Twisted Flux Tube into the Solar Atmosphere: Sunspot Rotations and the Formation of a Coronal Flux Rope. *Astrophys. J.* **697**, 1529-1542 (2009)

**Acknowledgements:** D.M., C.P. and B.R. welcome support from the American Air Force Office for Scientific Research (AFOSR): grant number FA8655-20-1-7032. S.L.G. and P.R. welcome support from the Italian MIUR-PRIN grant 2017APKP7T "Circumterrestrial environment: Impact of Sun-Earth Interaction" and by the Istituto Nazionale di Astrofisica (INAF). This research received funding from the European Union's Horizon 2020 Research and Innovation program under grant agreements No 824135 (SOLARNET) and No 729500 (PRE-EST). The simulation was performed using the ARCHIE-WeSt High Performance Computer (www.archie-west.ac.uk) based at the University of Strathclyde.

**Author contributions:** D.M. coordinated the project, co-developed the theoretical part of the work, performed the simulation, produced graphical output and drafted the manuscript; C.P. co-developed the theoretical part of the project, performed helicity and winding calculations and revised the manuscript; B.R. produced graphical output and revised the manuscript; P.R. performed helicity and winding calculations and revised the manuscript; S.L.G. selected the active regions for study and revised the manuscript

**Methods section:**

**Numerical simulation details:** The flux emergence simulation was performed using the open-source Lare3D code[31], which solves the equations of magnetohydrodynamics. The version we used includes an extended energy equation to allow for the modelling of convection, the details of which are described in a previous study[26]. The setup of the simulation is identical to that previous study, with the exception of the following parameters in the initial conditions: the flux tube has an axial field strength $B = 3.25 \times 10^3$ G and a uniform field line twist of 0.4 radians over a distance of 170 km.

**Helicity and winding flux calculations:** The helicity flux through the photospheric boundary $P$ is given by

$$\frac{dH}{dt} = -\frac{1}{2\pi} \int_{P \times P} \frac{d\theta(x, y)}{dt} B_z(x) B_z(y) \, d^2x \, d^2y,$$

where $B_z$ is the component of the magnetic field $\boldsymbol{B}$ orthogonal to $P$, $\boldsymbol{x}$ and $\boldsymbol{y}$ are position vectors in $P$ that mark the intersection of field lines with the plane, and $\theta(\boldsymbol{x}, \boldsymbol{y})$ is the angle made by the vector $\boldsymbol{x} - \boldsymbol{y}$ in $P$. The rate of change of this angle, measuring the rate of pairwise rotation of field lines, is given by

$$\frac{d\theta(\boldsymbol{x}, \boldsymbol{y})}{dt} = \boldsymbol{e}_z \cdot \frac{(\boldsymbol{x} - \boldsymbol{y})}{|\boldsymbol{x} - \boldsymbol{y}|^2} \times \left(\frac{d\boldsymbol{x}}{dt} - \frac{d\boldsymbol{y}}{dt}\right),$$

where $\boldsymbol{e}_z$ is the unit vector orthogonal to $P$. In the formula above, the motion of a point $\boldsymbol{x}$, representing the intersection of a field line with $P$, is given by the field line velocity $\boldsymbol{u}$[15,16],

$$\boldsymbol{u}(\boldsymbol{x}) = \frac{d\boldsymbol{x}}{dt} = \boldsymbol{v}_{||}(\boldsymbol{x}) - \frac{v_z(\boldsymbol{x})}{B_z(\boldsymbol{x})} \boldsymbol{B}_{||}(\boldsymbol{x}),$$

where $\boldsymbol{v}$ is the plasma velocity. The first term on the right-hand side is the in-plane velocity which braids the magnetic field in the atmosphere. The second term on the right-hand side is due to the emergence of the magnetic field, which causes apparent motion of the field line. The braiding contribution to helicity (and, similarly for the winding defined below) is found by setting $\boldsymbol{u} = \boldsymbol{v}_{\parallel}$. Similarly, the emergence contribution is found by setting $\boldsymbol{u} = -v_z \boldsymbol{B}_{\parallel}/B_z$.

The winding flux is given by

$$\frac{dL}{dt} = -\frac{1}{2\pi} \int_{P \times P} \frac{d\theta(\boldsymbol{x}, \boldsymbol{y})}{dt} \sigma_z(\boldsymbol{x}) \sigma_z(\boldsymbol{y}) \, d^2x \, d^2y,$$

where the $\boldsymbol{e}_z$-component of the magnetic field is replaced by the indicator function

$$\sigma_z(\boldsymbol{x}) = \begin{cases} 1 & \text{if } B_z(\boldsymbol{x}) > 0 \text{ and } |\boldsymbol{B}(\boldsymbol{x})| > \varepsilon, \\ -1 & \text{if } B_z(\boldsymbol{x}) < 0 \text{ and } |\boldsymbol{B}(\boldsymbol{x})| > \varepsilon, \\ 0 & \text{if } B_z(\boldsymbol{x}) = 0 \text{ or } |\boldsymbol{B}(\boldsymbol{x})| \leq \varepsilon, \end{cases}$$

where $\varepsilon$ is a field strength cut-off[25,26]. The above integrals were determined numerically using a standard trapezoidal method. In this work, the winding inputs from the observations are calculated with a cut-off of $\varepsilon = 50$G. Note that the 1-sigma error associated with SHARP data of the longitudinal magnetic field measurements is 10G. We have tested cut-offs up to $\varepsilon = 100$G, with no qualitative difference to our results.

As mentioned in the main text, plasma velocity $\boldsymbol{v}$ was obtained from observations using the DAVE4VM[28] method. The results presented were calculated with the version of the code written in Python but we have also checked the results using the version written in the Interactive Data Language (IDL). The SHARP vector magnetograms used for the presented results were downloaded using SunPy[32] from the Joint Science Operations Center (JSOC) database.

31. Arber, T.D., Longbottom, A.W., Gerrard, C.L., Milne, A.M. A Staggered Grid, Lagrangian-Eulerian Remap Code for 3-D MHD Simulations. *J. Comp. Phys.* **171**, 151-181 (2001)

32. The SunPy Community et al. The SunPy Project: Open Source Development and Status of the Version 1.0 Core Package. *Astrophys. J.* **890**, 68 (2020)

**Code availability:** We use a mixture of open-source and in-house codes for both simulations and observations. Since putting all these together is non-trivial, we invite interested parties to contact the authors directly for instructions on how to receive and use the codes.

**Data availability:** The observational data (winding and helicity ouptuts) used in this study will be made available. The simulation data will not be stored due to its prohibitively large size. Interested parties are welcome to contact us to learn how to recreate the simulation data.

---

Extended data figures:

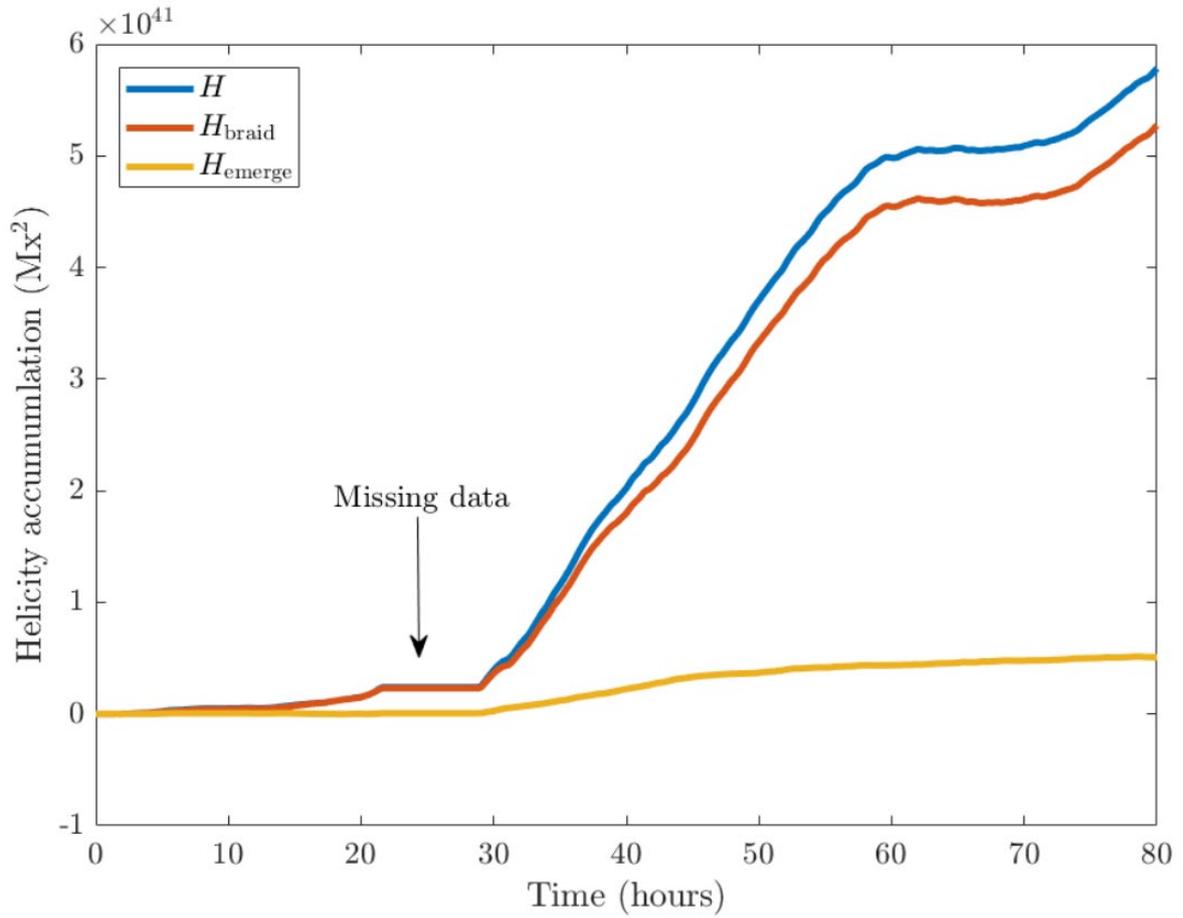

**Extended data figure 1: Helicity accumulation for AR11318.** The emergence accumulation $H_{emerge}$, the braiding accumulation $H_{braid}$, and the total helicity accumulation $H = H_{emerge} + H_{braid}$ are displayed.

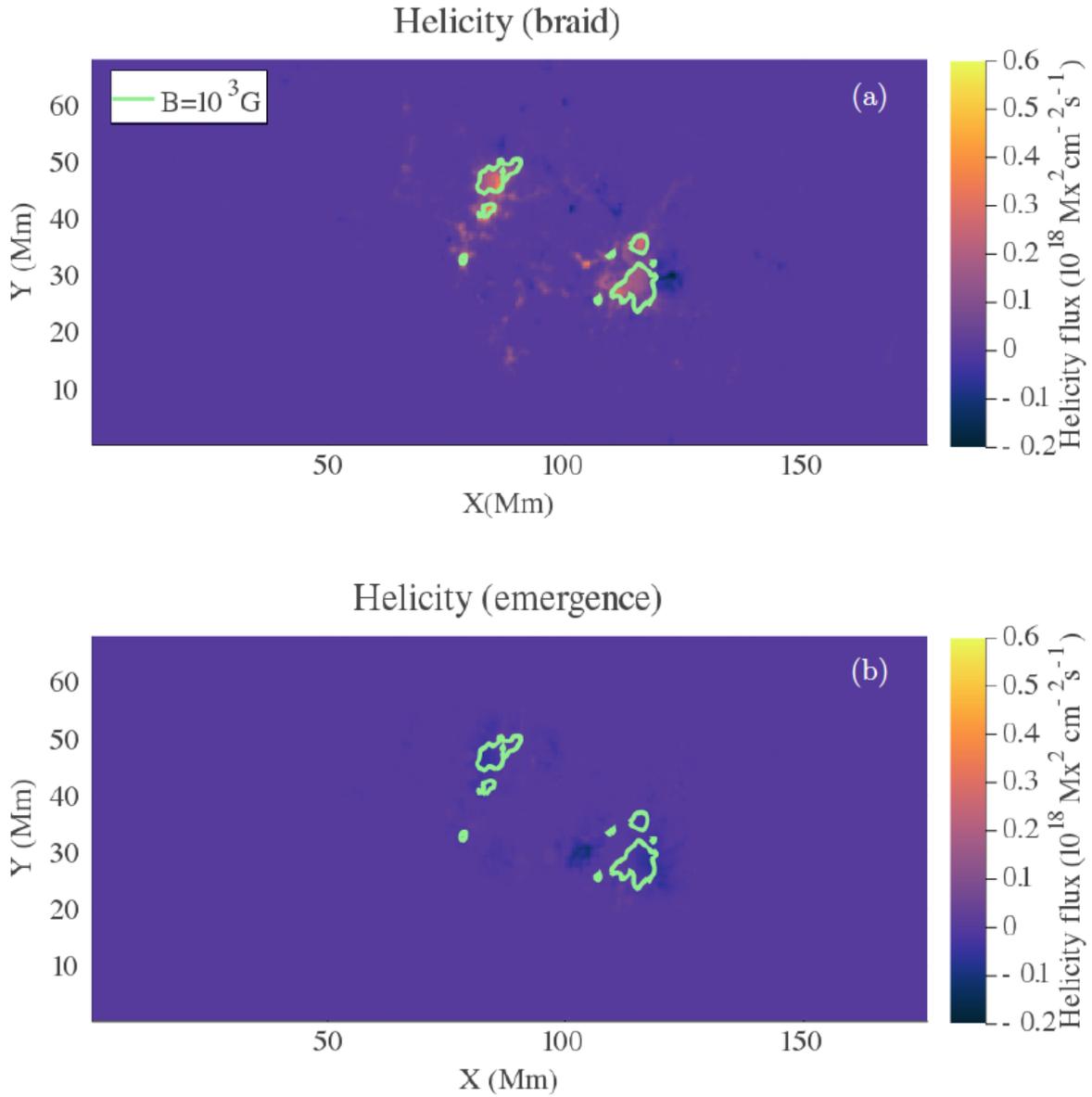

**Extended data figure 2: Helicity flux distributions.** Distributions of (a) $dH_{emerge}/dt$ and (b) $dH_{braid}/dt$ of AR11318 at $t = 42$ hours. The green contours indicate a field strength of $B_z = 1kG$ (and thus indicate the sunspot locations).

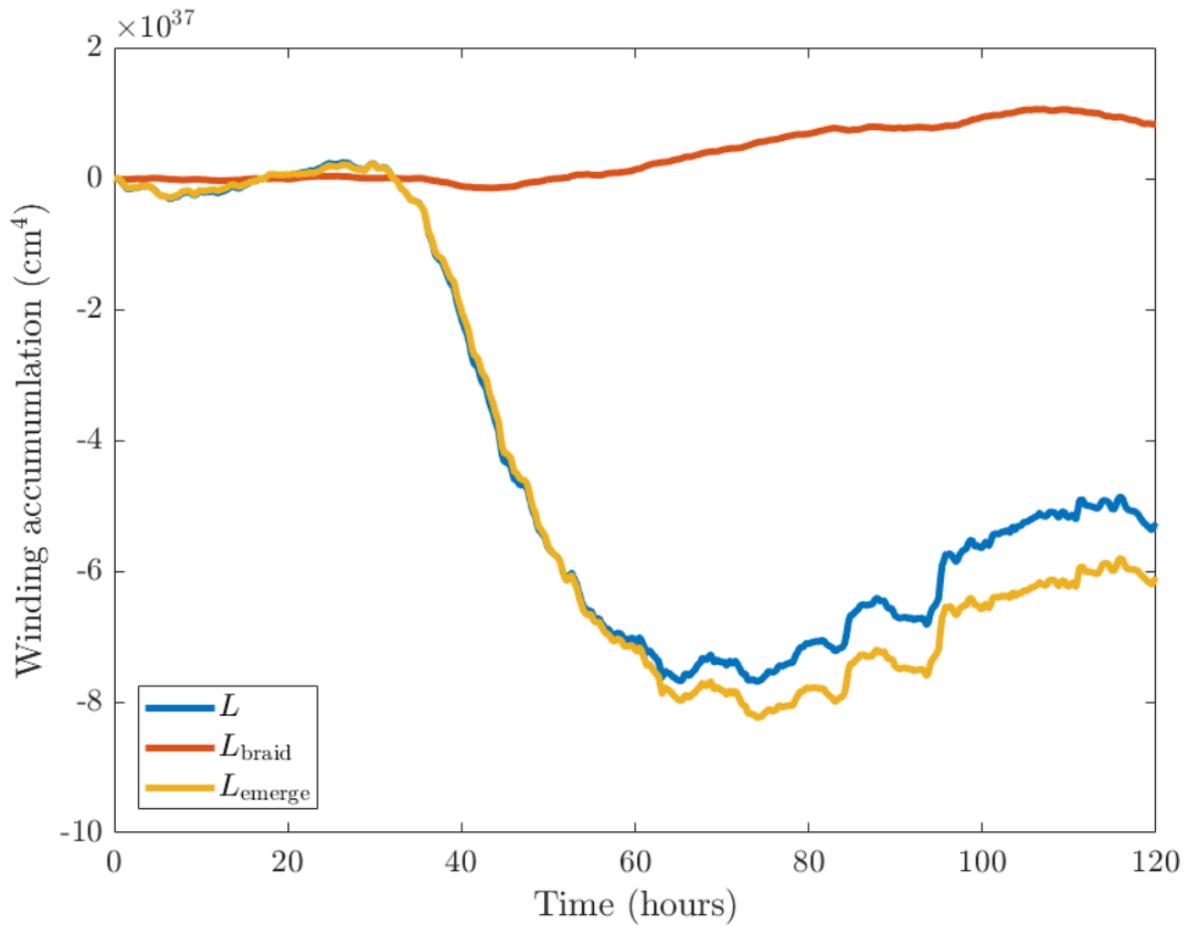

**Extended data figure 3: Winding accumulation of AR12203.** The emergence accumulation $L_{emerge}$, the braiding accumulation $L_{braid}$, and the total winding accumulation $L = L_{emerge} + L_{braid}$ are displayed.

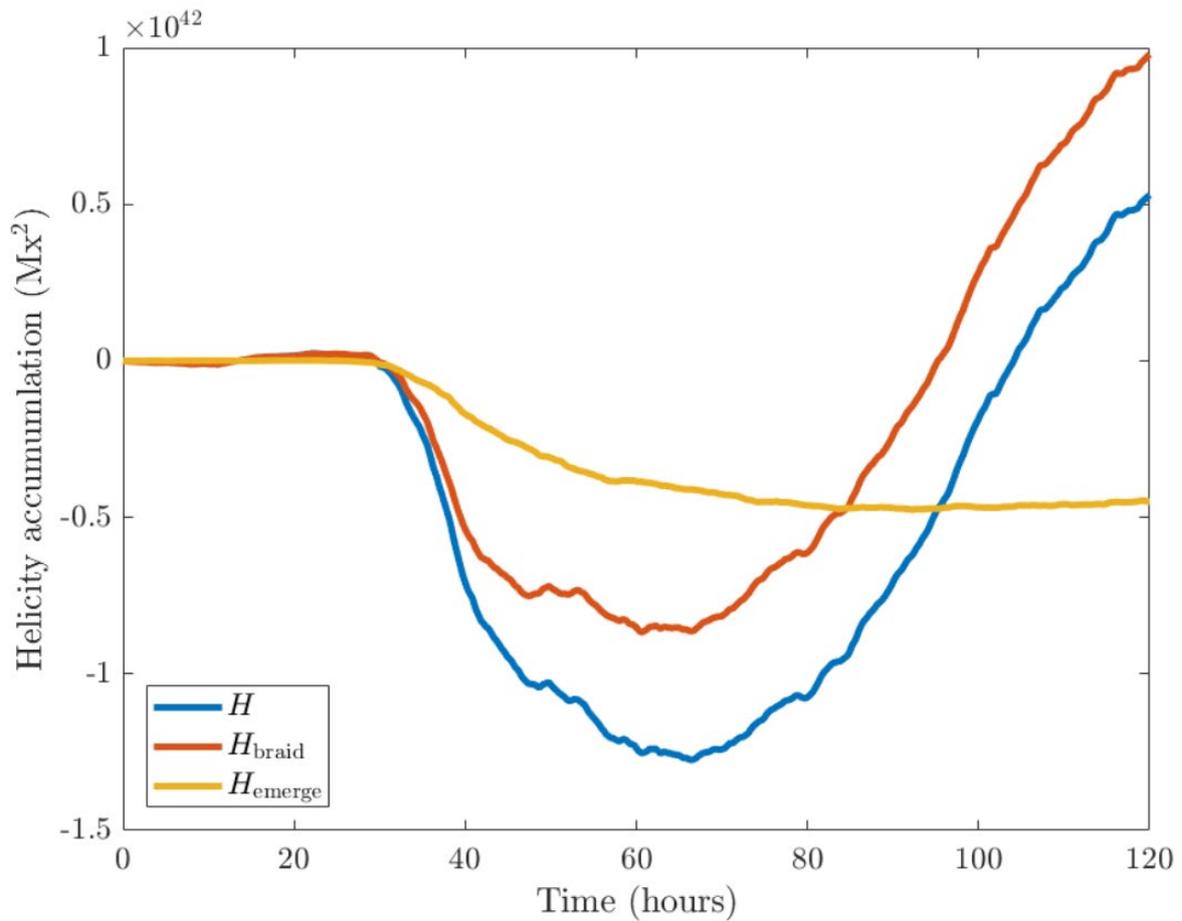

**Extended data figure 4: Helicity accumulation for AR12203.** The emergence accumulation $H_{emerge}$, the braiding accumulation $H_{braid}$, and the total helicity accumulation $H = H_{emerge} + H_{braid}$ are displayed. Notice the change in sign of the helicity accumulation when the winding accumulation (Extended data figure 3) is approximately constant by comparison.